\documentclass[a4paper,12pt,showpacs,preprintnumbers,showkeys
 amsmath,amssymb,
floatfix]{article}
\usepackage{graphicx}
\usepackage{infnprep}
\usepackage{dcolumn}% Align table columns on decima
\usepackage{hyperref}
\usepackage[top=2.5cm, bottom=3cm, left=3cm, right=3cm]{geometry}
\newcommand{\Header}{
  \resizebox{15cm}{!}{
  \begin{tabular}{rl}
  \includegraphics[width=5cm, trim={50 100 0 0}]{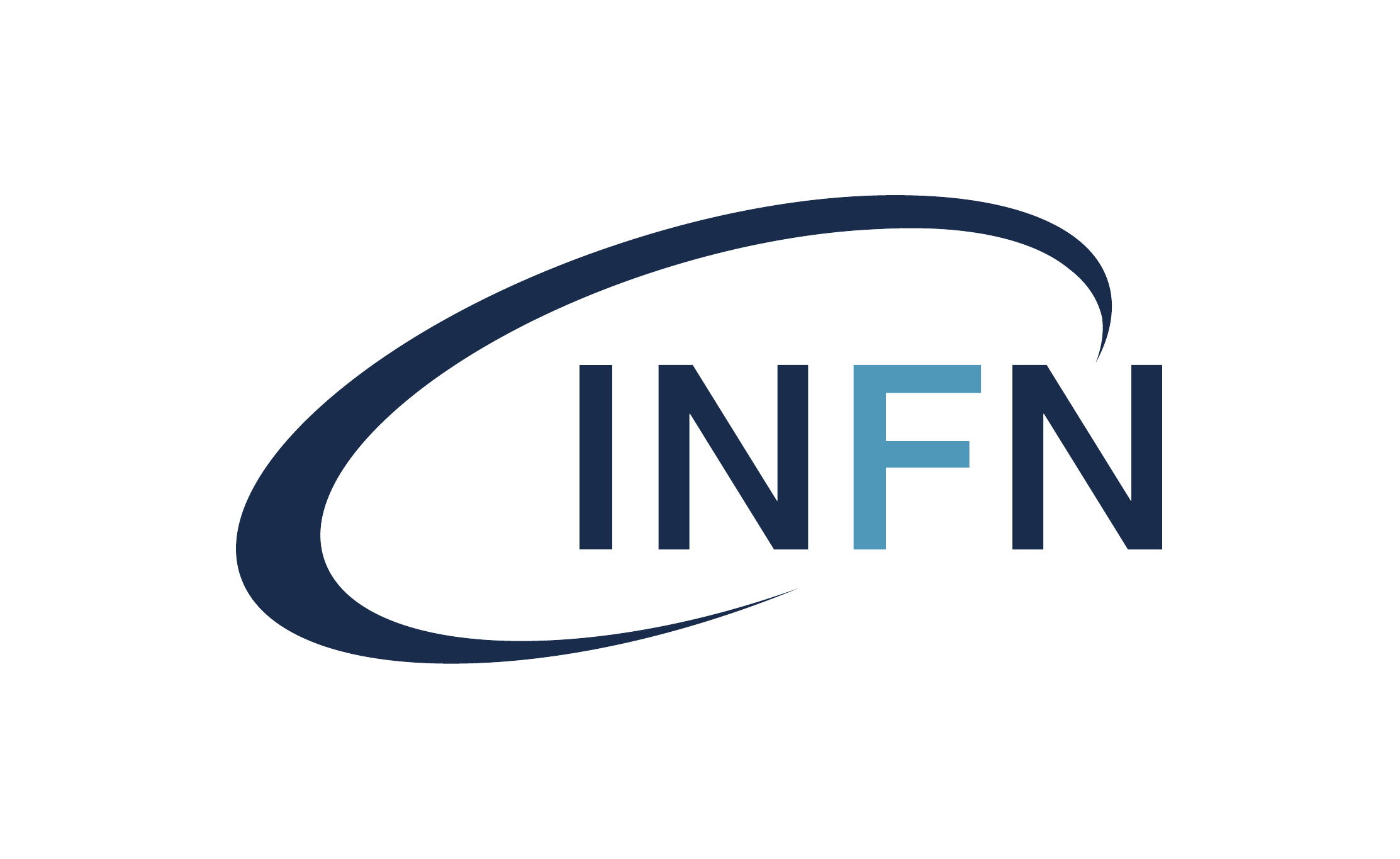} & {\LARGE\sffamily ISTITUTO NAZIONALE DI FISICA NUCLEARE}\\
      \\
  \end{tabular}
  }
    \renewcommand{\arraystretch}{1}
\vskip 0.5cm
\rule{15.0cm}{0.09mm}
\vskip 1.5cm
  \begin{flushright}
      {\underline{\bf INFN-19-08/LNF}}\\    % insert here the preprint number
      {\small\bf 6 May 2019} \\      % insert here the preprint Date
  \end{flushright}
}
\begin{document}
\begin{titlepage}
\title
  {\Header \large \bf Science Communication a New Frontier of Researcher's Job}

\author{
Giovanni Mazzitelli$^{1, 2}$\\
{\it $^{1)}$ INFN, Laboratori Nazionali di Frascati, Via E. Fermi, 40 - 00044 Frascati (RM), Italy} \\
{\it ${}^{2)}$ Associazione Frascati Scienza, Piazza Marconi 6 - 00044, Frascati (RM), Italy}
} 

\maketitle
\baselineskip=14pt

\begin{abstract}
In the world of communication, nobody can be out of the fray! Since many years science communication and more in general the ability of a researcher to communicate his/her work to founding agency, policy makers, entrepreneurs and public at large, starts to be a fundamental skill of the researcher’s job. This skill is needed and requested to access funds and successfully disseminate the research outcome, as well as to engage society in understanding science and its benefits. Moreover, due to the large decrease of research funds and of people starting scientific carrier, researchers must be in the front line to promote the scientific culture in order to invert the dreadful trend of last years. Where are we and where are we going to? We try to answer such questions introducing successful models that can be used without huge overloads for our job. 

This paper reports on the experience of one of the largest and oldest project in Europe of the Marie Sklodowska-Curie Actions European Researchers' Night and describes how this project followed the evolution in science communication.
\end{abstract}

\vspace*{\stretch{2}}
\begin{flushleft}
% insert here the PACS number 
\end{flushleft}
\begin{flushright}
  \vskip 3cm
\small\it Invited talk at  \\
2018 Channeling Conference Record, Ischia (NA), Italy \\
CERN Courier Volume 59, Number 3 - May/June 2019, pag 59
\end{flushright}
\end{titlepage}
\pagestyle{plain}
\setcounter{page}2
\baselineskip=17pt

\section{\label{sec:intro}Introduction}
Science communication is increasingly present in the researchers' job and career. No matter what is the science field, today the researcher has to communicate for many different reasons.

Firstly, we live in the age of social media, where any of one’s thought is of public domain, in the age of the \textit{google knowledge}, where everyone is an expert, in the age of fake news and fact checking, where everything become a talk show~~\cite{Quattrociocchi}. Last but not least, we live in the age in which nobody has \textit{free} to use communication, if not willing to be accused to hide something, especially when spending public's funds. This is particularly true for scientists. The scientists play a fundamental role in society, they benefit of a great influence and authoritativeness~\cite{observa}, which make them accountable towards the public. This suggests that science communication is not only a media to share your knowledge, to share the rational point of view that brought us to over 400 years human evolution, to educate new generation to technology innovation and more in general to a scientific approach or simply to bring young people to start scientific careers, but it's also a great \textit{responsibility for the scientists}.

Secondly, fundamental research and its follow up actions allowing the exploitation of its results and the development of technology and innovation, play a fundamental role for the solution of the societal challenges (health, food, safety, wellness, education, etc, etc)~\cite{eusc}. This obliges researchers not only to disseminate the results of their job (hopefully in the most open way), but also to actively involve entrepreneurs and policy makers in technology transfer chain and hopefully its funding. 

Thirdly, multidisciplinary and interdisciplinary evolution of the discoveries push researchers to be more and more accurate in the communication with their colleagues.

There exist nevertheless some risks. Although science communication is becoming an unavoidable necessity, to become an \textit{open scientist}, able and eager to communicate, such skill has not yet become “usual” and scientists are not prepared to it. Communication can distract from the research job and objectives, or, on the other side, can generate a doped \textit{marked} of science where the image provided is more important then the contents. Moreover, to be open, simple and accessible could also leave room for mystification with all the linked risks.

\section{\label{sec:evaluations}Experimenting the Science Communication Evolution}

In 2006 I was leading a small group of researchers from the Italian National Institute for Nuclear Physics (INFN) located close to Frascati town that, supported by the European Commission, realized one of the first European Researchers' Night (NIGHT)~\cite{EUNIGHT}: a one night-event that falls every last Friday of September to promote the researcher's figure and its work in Europe. The Commission started to support this project in the frame of the Marie Sklodowska Curie Action (MSCA) in 2005, under the PEOPLE specific programme; in such a context, a first public event was organized in Brussels by the Commission itself, aimed to enhance researchers’ public recognition, their role within the society and to stimulate youngsters and people at large to be involved in research and understand its impact in everyday life. 

Frascati is a small town 30~km far from Rome and is the epicenter of an area characterized by the presence of the most important Italian scientific institutions, the proximity of many universities, and a long list of science associations working to engage young people in the thrill of research and scientific discovery. Since the beginning, the Italian National Agency for New Technologies, Energy and Sustainable Economic Development (ENEA), the European Space Agency (ESA) and the National Institute for Astrophysics (INAF) joined the collaboration with INFN, as well as the Municipality of Frascati and the Cultural and Research Department of the Lazio Region, which co-funded the initiative.

\begin{figure}[htp]
\centering
\includegraphics[width=3.4in]{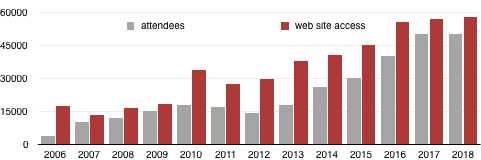}\DeclareGraphicsExtensions.
\caption{European Researchers’ Night attendees and web site access during the awareness campaign}
\label{fig:attendees}
\end{figure}

Today, the European Project BEES (BE a citizEn Scientist) is the most recent result  of a long evolution realized in Science communication by the Frascati's scientific area in the last thirteen years (Fig.~\ref{fig:attendees}). 

Thousands of researchers, citizens, public and private institutions worked together~\cite{Mazzitelli2016ICERIR}~\cite{Mazzitelli2018INTED} to change the public perception of science and of the infrastructures in the Frascati area and Lazio region, supported and pushed by the NIGHT project. Today, after thirteen editions, the project evolved by involving more than 60 scientific partners~\cite{FS} spread from the north to the south of Italy in 30 cities and is captivating more then 50.000 attendees whit an non negligible impact on the press and the territory~\cite{Mazzitelli2016ICERIL}. Moreover, the NIGHT has become a week-long event, linked to a lot of related events all over the year; it also triggered many institutions to develop science communication projects, adopting a new approach towards science communication and the relevance of the impact of science in society.

Analyzing the successive Frascati NIGHT’s projects allows a better understanding of the evolution of science communication methodology. The projects’ titles and objectives clearly underline the evolution of science communication that follow, and sometimes anticipate, a more general trend.
\begin{itemize}
\item 2006 – COME IN - NIGHT-2016-FP6-GA-044837: let's share thousand emotions together; the focus was to bring public to visit the laboratories and the research infrastructures in the area.
\item 2007 – AGORA - NIGHT-2007-FP7-GA-200202̀: enjoy beyond/with research/ers; the focus was to contaminate the cities in the Frascti research area.
\item 2008 – EOS - NIGHT-2008-FP7-GA-228619: Eyes On Researchers; the focus was to increase the knowledge of the scientific wealth of Frascati area
\item 2009 – SAY - NIGHT-2009-FP7-GA-244954: Scientist Around Youth; the focus was to engage young people to undertake the scientific careers.
\item 2010 – BEST - NIGHT-2010-FP7-GA-265743: Being a European Scientist Today; the focus was to show the large collaboration at European level
\item 2011 – BRAIN - NIGHT-2011-FP7-GA-287442: Be in contact with Research And its Institutions Network for a night; the focus was to give the image of a science without borders.
\item 2012 – RESPEcT - NIGHT-2012-FP7-GA-316436: RESearchers - Pure Energy from Tip to Toe; the focus was to enhance the relation with the researcher.
\item 2013 – TRAiL - NIGHT-2013-FP7-GA-609662: Taste the ReseArchers' Life; the focus was to show the researcher life.
\item 2014/15 – DREAMS - H2020-MSCA-NIGHT-2014/15-GA-633230: focus on Sustainability and Responsibility of Researchers and Science
\item 2016/17 – MADE IN SCIENCE - H2020-MSCA-NIGHT-2016/17-GA-722952: focus on the Science as creating effective benefits for European citizens
\item 2017/18 - BEES - H2020-MSCA-NIGHT-2018/19-GA-818728: focus on participated and collaborated science.
\end{itemize}

Detailed impact assessment reports and data collected through the survey conducted during 13 years of activities are available in reference~\cite{FSreports}.

The Frascati NIGHT started in 2006, when people began benefiting from digitization and new media communication, the economy was growing up and Europe was seen as an opportunity as well as a community. At the beginning the projects aimed to bring people to know, educate and understand science. The researchers started to open their laboratories and research infrastructures, to show their job in the most comprehensible way and with a view to increasing the scientific literacy of public and fill their \textit{deficit} of knowledge. Then they started to meet people in public spaces, such as squares, pubs, etc, etc. \textit{playing} with public, trying to create a direct \textit{dialogue}, to explain how public money where spent, and how much researchers are responsible concerning their job. 

Those were the years in which the socio-economic crisis started, in which Europe as well as the scientists were imagined as a safe refuge, those were the years when the H2020 program started, addressing economic growth, and probably not fully computed. Then, those were the years in which the word \textit{innovation} started to substitute \textit{scientific progress} and \textit{discovery}, somehow infringing the pureness of science. Those were the years (not finished yet) of the Sustainability and of the Responsible Research and Innovation (RRI), in which the dialogue with the researchers might have been insufficient to keep the science flag flying.

That bring us to the last years when two biannual projects, MADE IN SCIENCE and BEES, try to underlain a different vision of science and of the methodology of communication. “MADE IN SCIENCE” (2016-2017) was supposed to represent the \textit{trademark of research}, which shows the strength of researcher's job. The aim of the project was to communicate to society the importance of the science \textit{production chain} in terms of quality, identity, creativity, security guarantee, transnationality, know-how and responsibility. In this chain, that starts from fundamental research and ends with social benefits, no one is excluded and must take part in the decision process and whether possible in the research itself. In fact, BEES (2018-2019), brings citizens to \textit{become scientists} involved in the discovery process, also showing how long it takes and how it can be tough and frustrating. This brings us to develop a set of \textit{citizen science mini project}~\cite{WCS}, especially addressing young people.  

Both projects MADE IN SCIENCE and BEES were designed to following the most recent steps in science communication, based on the \textit{participation} model. The participation is then the basis of what is currently called \textit{public engagement}~\cite{PPE} in which people are involved from the beginning of the science communication project addressing and co-designing how to bring a message and if possible in the research itself, as in the citizens science. This marks a revolution in researcher’s communication, no more as main actor like in \textit{deficit} and \textit{dialogue} model, but as a facilitator of the learning process with a role: the \textit{expert} one.

\begin{figure}[htp]
\centering
\includegraphics[width=3.4in]{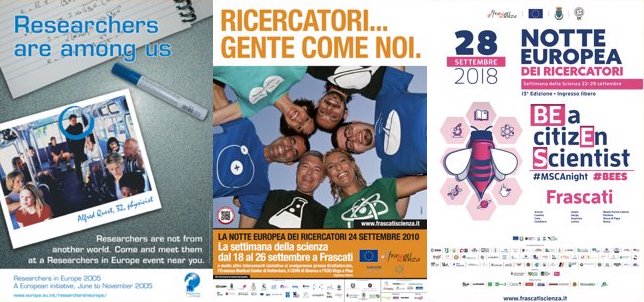}\DeclareGraphicsExtensions.
\caption{Evolution of the Researchers' night posters: (left) 2005 - Researchers are among us; (center) 2010 - Researchers are people like us; (right) 2018 - Be a Citizen Scientist}
\label{fig:ads}
\end{figure}

It's interesting to notice how in the mean time the way to communicate completely changed since 2006 (see Fig.\ref{fig:ads}): from large, expensive, posters on public transport broadcasting the image of researchers' faces to a logo that should represents a community where everyone finds their role. \textit{A science of everybody and for everyone}.

The projects also got enhanced since 2010 when communication professionals started to be part of the project team, strongly increasing the impact of the awareness campaign and press outcome. In the meantime, the start of social media helped to retaining the interested audience and to enlarge the targets with an inexpensive, but well-addressed message. Social media also opened the possibility to easily develop audio-video contents, allowing to create contents with a double aim: be aware about the project and to communicate a science message. This is perfectly illustrated in the so-called \textit{science pills}, funny videos with many tens of millions of visualizations, awarded by the scientific community, in the short movie international festivals and as the best non-profit sustainable advertising in 2017~\cite{FSspot}

\section{\label{sec:issue}Open Issues}

Such evolution of science communication can't nevertheless be evaluated only as positive. There are many outcomes that probably resulted from this recent evolution pressing the science communication debate. There are many examples of huge and actual problems in science communication: the explosion of concerns about science (vaccines, autism, GMO, homeopathy, etc)~\cite{observa}; the risks due to the wellness, in which people live like the inductivist turkey~\cite{russel}, forgetting the past and avoiding science results in the name of the return to a more \textit{natural} life; the funds coercion limiting the freedom of research~\cite{Mazzitelli:2016mela}, strongly addressing how and why to do research; the exploitation of science results (positive or negative) to argue \textit{my} theses~\cite{sloman2018illusione}, in support to conspiracy theories, democracy, populism; the limits of science projects’ evaluation (e.g. the benefit provided by the funding agency for knowledge and technology transfer of which science communication is now part) and of the Journal Impact Factor~\cite{JIF}, sometimes doping the \textit{market of publications}; the predatory publishers~\cite{predatory} that exploit the researchers’ needs in enhancing their image by publishing fake papers without peer review and fake conferences behind rewarded.

Last but not least, some strong bias still remain in both scientists and audiences, limiting the communication. The first one, and probably the hardest, is the \textit{stereotype bias}: Are you a nerd, or do you feel a nerd? Often scientists refer to themselves as a category that can’t be understood by the society, consequently limiting their capacity to interact with the public; on the other hand, scientists are sometimes real nerds,  seen by the public as nerds. This is true for all the job categories, but in the case of scientists this strongly conditions their ability to communicate. 

\textit{Age, gender and technological bias}~\cite{eurostat} are still playing a fundamental role, especially in the most developed European countries. The young people may understand much easily science and technology while women still seem not to have full access to scientific careers and to the exploitation of technology. Moreover, it’s a \textit{social bias}~\cite{eurobarometer}, affecting both scientists and audience and conditioning the rationality of the message. Scientists are anyway human, whit all their inherent bias about life, beliefs etc, etc. conditioning not only both receiver but also the sender of the message.

Finally, a \textit{methodological bias}, only partially solved by the participating model, is still existing. Our brain doesn't contain more than one gigabyte. Nothing compared to how much information a computer and the network can make available. Yet we insist on being competent, unique and much smarter than others. We continually overestimate ourselves. In reality we are very ignorant and live in \textit{the knowledge illusion}~\cite{sloman2018illusione} and only through access to the community of knowledge can we fill our gaps and progress socially and scientifically. It is beautiful that this human limit is overcome only by the causal process that in our brain is triggered through a group intelligence, and that this is strongly true also for scientists and their interaction with the public.

\section{\label{sec:colclusion}Conclusions}
In conclusion, science communication has achieved huge progress during the most recent years. This results from the increasing engagement of researchers and their innovative way to communicate. Although, the transition from a \textit{deficit model} to a \textit{participated model} is not belonging only to science, but is common in education and in democratic society, this transition is not yet completed in science, probably due to the strong bias that are still among the researchers and the audience. This limits may be the origin of the current public debate about science that the researchers have, for first, the responsibility to overcome.

Obviously, issues connected to the failure science communication on themes such as environmental and climate change and to the innumerable ethical problems of some scientific disciplines, go far beyond the experiences of organizing scientific communication events and without a doubt add to the difficulties encountered in science communication.

\section{\label{sec:ack}acknowledgement} 
I would like to strongly thank Sultan Dabagov, who give me the opportunity to underline this new and challenge part of the researcher' job and all the volunteers of Frascati Scienza team, true actors of the change as well as Colette Renier (EU Commission Project Officer) who drove us during all these years.

\bibliographystyle{unsrt}
\bibliography{main}

\begin{thebibliography}{10}

\bibitem{Quattrociocchi}
M.~Del~Vicario et~al.
\newblock The spreading of misinformation online.
\newblock {\em Proceedings of the National Academy of Sciences},
  113(3):554--559, 2016.

\bibitem{observa}
G.~Pellegrini.
\newblock {\em Annuario scienza tecnologia e societ{\`a} (2018)}.
\newblock Fuori collana. Il Mulino, 2018.

\bibitem{eusc}
{Societal Challenges}.
\newblock
  \url{https://ec.europa.eu/programmes/horizon2020/en/h2020-section/societal-challenges},
  2018.

\bibitem{EUNIGHT}
{European Commission}.
\newblock {European Researchers’ Night MSCA-NIGHT}.
\newblock \url{http://ec.europa.eu/research/researchersnight}, 2018.

\bibitem{Mazzitelli2016ICERIR}
G.~Mazzitelli et~al.
\newblock {RESULTS FROM IMPACT ASSESSMENT ON SOCIETY AND SCIENTISTS OF FRASCATI
  SCIENZA EUROPEAN RESEARCHERS’ NIGHTS IN YEARS 2006 – 2015}.
\newblock 9th annual International Conference of Education, Research and
  Innovation, pages 3356--3365. IATED, 14-16 November, 2016 2016.

\bibitem{Mazzitelli2018INTED}
G.~Mazzitelli et~al.
\newblock {12 YEARS OF DATA, RESULTS AND EXPERIENCES IN THE EUROPEAN
  RESEARCHERS’ NIGHT PROJECT}.
\newblock 12th International Technology, Education and Development Conference,
  pages 1772--1780. IATED, 5-7 March, 2018 2018.

\bibitem{FS}
{Associazione Frascati Scienza}.
\newblock \url{http://www.frascatiscienza.it/}, 2008.

\bibitem{Mazzitelli2016ICERIL}
G.~Mazzitelli et~al.
\newblock {UNIVERSITY, INDUSTRY AND RESEARCH COOPERATION: THE LAZIO PULSE
  INITIATIVE}.
\newblock 9th annual International Conference of Education, Research and
  Innovation, pages 3808--3814. IATED, 14-16 November, 2016 2016.

\bibitem{FSreports}
{Associazione Frascati Scienza}.
\newblock Impact assessment reports.
\newblock \url{https://www.frascatiscienza.it/report-impact/}, 2018.

\bibitem{WCS}
{Citizen Science}.
\newblock \url{https://en.wikipedia.org/wiki/Citizen_science}, 2018.

\bibitem{PPE}
{Sense about Science}.
\newblock Public engagement: a practical guide.
\newblock
  \url{http://senseaboutscience.org/activities/public-engagement-guide/}, 2017.

\bibitem{FSspot}
{Associazione Frascati Scienza}.
\newblock Xiv edition of price san bernardino, october 2016; price bayer “xii
  edizione di cortinametraggio”, cortina, italy, march 2017; awarded at 13rd
  edition of cinema italian style, los angeles 16-31 november 2017.
\newblock \url{https://www.youtube.com/watch?v=Ppj_3MFBKmQ&t=2s}, 2017.

\bibitem{russel}
Bertrand Russell.
\newblock {\em {I Problemi della Filosofia}}.
\newblock Feltrinelli, 1988.

\bibitem{Mazzitelli:2016mela}
G.~Mazzitelli.
\newblock {La scienza e il feedback del passato}.
\newblock {\em La Mela di Newton}, 2016.

\bibitem{sloman2018illusione}
S.~Sloman and P.~Fernbach.
\newblock {\em {The Knowledge Illusion: Why We Never Think Alone}}.
\newblock Expert Thinking Series. Pan Macmillan, 2017.

\bibitem{JIF}
{Journal Impact Factor (JIF)}.
\newblock \url{https://en.wikipedia.org/wiki/Impact_factor}, 2018.

\bibitem{predatory}
Predatory open-access publishing.
\newblock \url{https://en.wikipedia.org/wiki/Predatory_open-access_publishing},
  2018.

\bibitem{eurostat}
{Eurostat}.
\newblock About science and technology.
\newblock
  \url{https://ec.europa.eu/eurostat/statistics-explained/index.php?title=Science_and_technology},
  2018.

\bibitem{eurobarometer}
{Eurobarometer}.
\newblock {PUBLIC OPINION ON FUTURE INNOVATIONS, SCIENCE AND TECHNOLOGY,
  Aggregate Report June 2015}.
\newblock
  \url{http://ec.europa.eu/commfrontoffice/publicopinion/index.cfm/ResultDoc/download/DocumentKy/65618},
  2015.

\end{thebibliography}

\end{document}